\documentclass[prl, final, twocolumn, showpacs]{revtex4}

\usepackage{amsmath}
\usepackage{graphicx}
\usepackage{amsfonts}
\usepackage{amssymb}
\usepackage{subfigure}
\usepackage[]{silence}
\usepackage{titlesec}
\usepackage{cases}
\usepackage{siunitx}
\usepackage{braket}

\newcommand{\beginsupplement}{%
    \setcounter{section}{0}
    \renewcommand{\thesection}{S\Roman{section}}%
    \setcounter{subsection}{0}
    \renewcommand{\thesubsection}{\Alph{subsection}}%
    \setcounter{subsubsection}{0}
    \renewcommand{\thesubsubsection}{S\Alph{subsubsection}}%
    \titleformat{\subsubsection}[block]{\bfseries\centering}{\thesubsubsection.}{1em}{}
    \setcounter{table}{0}
    \renewcommand{\thetable}{S\Roman{table}}%
    \setcounter{figure}{0}
    \renewcommand{\thefigure}{S\arabic{figure}}%
    \setcounter{equation}{0}
    \renewcommand{\theequation}{S\arabic{equation}}%
    }

\newcommand{\subs}[1]
{ 
	\mbox{\scriptsize{#1}}
}

\begin{document}

\title{Multiphoton Quantum Logic Gates for Superconducting Resonators with Tunable Nonlinear Interaction}

\author{Frederick W.~Strauch$^1$}
\author{Matteo Mariantoni$^{2,3}$}
\affiliation{$^1$Williams College, Williamstown, MA 01267, USA \\
$^2$Institute for Quantum Computing, University of Waterloo, 200 University Avenue West, Waterloo, Ontario N2L 3G1, Canada \\
$^3$Department of Physics and Astronomy, University of Waterloo,
200 University Avenue West, Waterloo, Ontario N2L 3G1, Canada}

\date{\today}

\begin{abstract}
We propose a tunable nonlinear interaction for the implementation of quantum logic operations on pairs of superconducting resonators, where the two-resonator interaction is mediated by a transmon quantum bit (qubit). This interaction is characterized by a high on-to-off coupling ratio and allows for fast qubit-type and $d$-level system (qudit)-type operations for quantum information processing with multiphoton cavity states. We present analytical and numerical calculations showing that these operations can be performed with practically unit fidelity in absence of any dissipative phenomena, whereas physical two-photon two-resonator operations can be realized with a fidelity of~\SI{99.9}{\percent} in presence of qubit and resonator decoherence. The resonator-qubit-resonator system proposed in this Letter can be implemented using available planar or three-dimensional microwave technology.
\end{abstract}


\maketitle

\textit{Introduction.}\textemdash Major growth in the field of superconducting quantum circuits~\cite{Schoelkopf2008, Clarke2008, Wendin2017} has led to the possibility of controlling quantum microwave photons in superconducting resonators (or cavities) and transmission lines~\cite{Gu2017}. This includes the preparation of arbitrary photon states in one resonator~\cite{Hofheinz2009} and NOON states in two planar resonators~\cite{Wang11}, as well as Schr{\"o}dinger cat states in one~\cite{Vlastakis2013} and two three-dimensional cavities~\cite{Wang2016}. However, taking full advantage of superconducting resonators for quantum computing applications requires a universal quantum logic gate on their microwave photonic states. Progress toward this goal includes theoretical work studying single resonators as quantum bits (qubits)~\cite{Adhikari2013}, or $d$-level systems (qudits)~\cite{Strauch2011, Strauch2012, Krastanov2015}, as well as recent experimental work demonstrating effective two-qubit interactions between a qubit and the modes of a resonator~\cite{Naik2017} and the unitary control of two or more states of cavity~\cite{Heeres2015, Heeres2017}. Most recently, a specific multiphoton logic gate has been performed by a sequence of interactions of two cavities~\cite{Rosenblum2018}. Continued progress toward full unitary control of multiphoton states in multiple resonators constitutes an important and outstanding problem for quantum microwave photonics.

The ideal resonator-resonator interaction is represented by a tunable cross-Kerr Hamiltonian
\begin{equation}
\mathcal{H}_{\subs{cross-Kerr}} = \chi n_{\textrm{a}} n_{\textrm{b}},
\label{eq1}
\end{equation}
where~$\chi$ is a tunable coupling coefficient and $n_{\textrm{a}}$ and $n_{\textrm{b}}$ are the photon numbers in resonators~$R_{\textrm{a}}$ and $R_{\textrm{b}}$, respectively. This interaction is a powerful primitive for controlling multiphoton quantum states. For example, by turning on the interaction in Eq.~(\ref{eq1}) for a time~$t = \pi / \chi$, a product of coherent states evolves to an entangled state~\cite{Liu2017}. This can be used to implement a logic gate on the Schr{\"o}dinger's cat states
\begin{equation}
\ket{\mathcal{C}_{\alpha}^{\pm}} = \mathcal{N} \left( \ket{\alpha} \pm \ket{-\alpha} \right) \, ,
\end{equation}
where~$\mathcal{N}$ is a normalization constant for the superposition of the coherent states~$\ket{\alpha}$ and $\ket{-\alpha}$. These states have been proposed for hardware efficient quantum error correction~\cite{Leghtas2013} in three-dimensional cavities. By encoding quantum information in such multiphoton states, errors due to photon loss can be detected and corrected similar fashion to traditional quantum error correction~\cite{Gottesman2010}.  For two-cavity cat states, it can be shown that the cross-Kerr interaction generates a \textsc{controlled-phase} gate of the form $\ket{\mathcal{C}_{\alpha}^-} \ket{\mathcal{C}_{\beta}^-} \to -\ket{\mathcal{C}_{\alpha}^-} \ket{\mathcal{C}_{\beta}^-}$, where cavities~$R_{\textrm{a}}$ and $R_{\textrm{b}}$ are populated with cat states of complex amplitude~$\alpha$ and $\beta$, respectively, and with the other two-cavity states ($++, +-, \ \mbox{ and} \ -+$) unaffected. This can be seen by direct calculation or as a consequence of the fact that $\ket{\mathcal{C}_{\alpha}^{\pm}}$ are superpositions of even (for~$+$) and odd (for~$-$) numbers of photons.

The cross-Kerr interaction can also be used to implement a \textsc{controlled-phase} gate on the multiphoton states $\ket{0}_{\textrm{L}} = \ket{2}$ and $\ket{1}_{\textrm{L}} = (\ket{0} + \ket{4}) / \sqrt{2}$ used in the binomial state encoding scheme~\cite{Michael2016}. By turning on the cross-Kerr interaction for a time~$\pi/ (4 \chi)$, we find $\ket{0}_{\textrm{L}} \ket{0}_{\textrm{L}} \to -\ket{0}_{\textrm{L}} \ket{0}_{\textrm{L}}$, with the other states ($01, 10, \ \mbox{and} \ 11$) unaffected. An alternative gate on these states has been implemented in a recent experiment~\cite{Rosenblum2018} by the sequential microwave-driven coupling of the cavity states through the second excited state of a transmon qubit~\cite{Koch07}. The cross-Kerr implementation would also work for the multiphoton states studied in the work of Ref.~\cite{Heeres2017}.

\begin{figure}[b]
\includegraphics[width=3 in]{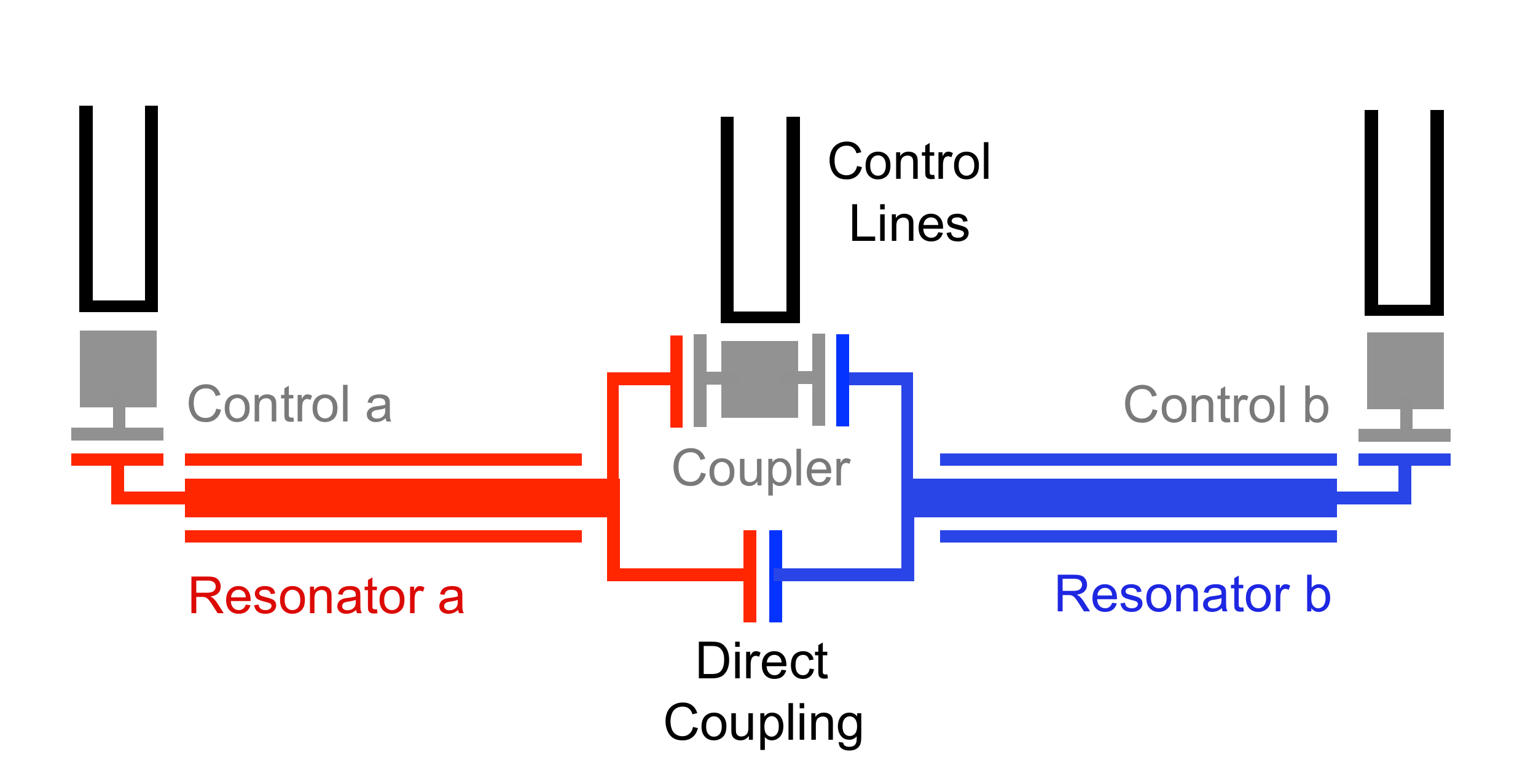}
\caption{Schematic representation of the resonator-qubit-resonator circuit used for tunable coupling and logic gates between two resonators. The resonators~$R_{\textrm{a}}$ and $R_{\textrm{b}}$ are capacitively coupled to a tunable qubit and directly to each other. Two auxiliary qubits, $C_{\textrm{a}}$ and $C_{\textrm{b}}$, make it possible to control~$R_{\textrm{a}}$ and $R_{\textrm{b}}$ individually.}
	\label{circuit}
\end{figure}

In this Letter, we explore how a simple resonator-qubit-resonator~(RQR) system, based on a tunable superconducting qubit or \textit{coupler}, such as the Xmon transmon qubit~\cite{Barends2013}, can be used to generate the nonlinear interaction necessary to implement the logic gates described above. An example of such a system is shown in Fig.~\ref{circuit}, where each cavity is depicted as a coplanar waveguide resonator with its own control qubit (for one-resonator operations~\cite{Heeres2015, Heeres2017}); the resonators are capacitively coupled to a tunable qubit and directly to each other. There are two main challenges to be faced when using such a system for multiphoton gates. First, a large inter-cavity coupling requires a small detuning between the qubit coupler and the resonators; this results in significant mixing of the qubit and photon states. Second, the cross-Kerr interaction is one of many terms that influence the overall dynamics; engineering the desired operation requires a careful analysis of the system. We address these challenges through analytical and numerical simulations of the RQR system.

We first introduce the nonlinear interactions achievable with the RQR system and study their properties. The linear interaction provided by such a scheme has been first studied theoretically in the work of Ref.~\cite{Mariantoni08} and more recently demonstrated experimentally~\cite{Baust2015}.  Nonlinear interactions have been considered using an idealized coupling scheme~\cite{Liu2017}, as well as through perturbation theory~\cite{Elliott2017}. We then demonstrate, by means of analytical calculations and numerical simulations, how to dynamically control the RQR system to provide a tunable strong coupling and implement fast two-qubit gates on the two resonators, where the resonator states are used as the computational basis. Finally, we show how to implement multiphoton operations using a dynamical sequence to synthesize the cross-Kerr interaction. Our results demonstrate that the RQR system can be used to achieve full unitary control of multiphoton quantum states of two coupled resonators.

\textit{Coupling Scheme.}\textemdash The basic features of the RQR system can be modeled by a two-mode Jaynes-Cummings-type Hamiltonian ($\hbar = 1$):
\begin{eqnarray}
\mathcal{H} &=& \omega_{\textrm{a}} a^{\dagger}a  + \omega_{\textrm{q}} \sigma^{\dagger} \sigma + \omega_{\textrm{b}} b^{\dagger} b + \ g_{\textrm{ab}} \left(a^{\dagger} b + b^{\dagger} a \right)  \nonumber \\
& & + g_{\textrm{a}} \left(a^{\dagger} \sigma + a \sigma^{\dagger} \right) + g_{\textrm{b}} \left( b^{\dagger} \sigma + b \sigma^{\dagger} \right).
\label{modelx}
 \end{eqnarray}
Here the resonant modes of $R_{\textrm{a}}$ and $R_{\textrm{b}}$ are represented by bosonic destruction and creation operators~$a$ and $b$ and $a^{\dagger}$ and $b^{\dagger}$ and have angular frequencies~$\omega_{\textrm{a}}$ and $\omega_{\textrm{b}}$, respectively. These modes interact with the qubit with coupling coefficients~$g_{\textrm{a}}$ and $g_{\textrm{b}}$, respectively, and with each other with coupling coefficient~$g_{\textrm{{ab}}}$. The qubit is represented by lowering and raising operators~$\sigma$ and $\sigma^{\dagger}$ and has a tunable angular frequency~$\omega_{\textrm{q}}$, which leads to an effective interaction between the two resonators.

To better understand the two-resonator Hamiltonian of Eq.~(\ref{modelx}), we first consider the case~$\omega_{\textrm{a}} = \omega_{\textrm{b}} = \omega$, $g_{\textrm{a}} = g_{\textrm{b}} = g$, and $g_{\textrm{ab}} = 0$. By introducing the normal-mode operators~$c_{\pm} = ( a \pm b ) / \sqrt{2}$, Eq.~(\ref{modelx}) becomes 
\begin{equation}
\mathcal{H} = \omega c_{-}^{\dagger} c_{-} + \omega_{\textrm{q}} \sigma^{\dagger} \sigma + \omega c_{+}^{\dagger} c_{+} + \sqrt{2} g \left( c_{+}^{\dagger} \sigma + c_{+} \sigma^{\dagger} \right).
\label{modely}
\end{equation}
It is possible to diagonalize Eq.~(\ref{modely}) exactly, from which we obtain the effective Hamiltonian
\begin{equation}
\mathcal{H}_{\subs{eff}} = \omega c_{-}^{\dagger} c_{-} 
+ \omega  c_{+}^{\dagger} c_{+} + \frac{1}{2} \left( \Delta - \sqrt{\Delta^2 + 8 g^2 (c_+^{\dagger} c_-)} \right),
\end{equation}
where~$\Delta = ( \omega_{\textrm{q}} - \omega )$ and we are using the dressed basis \cite{Boissonneault2009}, in which the coupler qubit is in its ground state. For~$\Delta \gg g$, we find
\begin{equation}
\mathcal{H}_{\subs{eff}} \simeq \omega c_{-}^{\dagger} c_{-} 
+ \left( \omega - 2 \frac{g^2}{\Delta} \right) c_{+}^{\dagger} c_{+} + 4 \frac{g^4}{\Delta^3} \left( c_{+}^{\dagger} c_{+} \right)^2. \end{equation}
The interaction with the qubit induces an ac~Stark shift and a Kerr interaction for the ``$+$'' mode (with ``$-$'' mode unaffected).

Returning to the original mode operators, we have
\begin{eqnarray}
\mathcal{H}_{\subs{eff}} &\simeq& \left( \omega - \frac{g^2}{\Delta} + \frac{g^4}{\Delta^3}  \right) a^{\dagger} a + \left(\omega - \frac{g^2}{\Delta} + \frac{g^4}{\Delta^3}  \right) b^{\dagger} b \nonumber \\
& & + \frac{g^4}{\Delta^3} (a^{\dagger} a)^2 + \frac{g^4}{\Delta^3} (b^{\dagger} b)^2 + 4 \frac{g^4}{\Delta^3} (a^{\dagger} a) (b^{\dagger} b) \nonumber \\
& & - \frac{g^2}{\Delta} \left( a^{\dagger} b + b^{\dagger} a \right) + 2 \frac{g^4}{\Delta^3} \left( a^{\dagger} a + b^{\dagger} b \right) \left( a^{\dagger} b + b^{\dagger} a \right) \nonumber \\
& & + \frac{g^4}{\Delta^3} \left[ (a^{\dagger})^2 b^2 + (b^{\dagger})^2 a^2 \right].
\end{eqnarray}
The coupling to the qubit has induced: frequency shifts and self-Kerr effects for each mode; a cross-Kerr interaction; a linear, a number-state-dependent, and a two-photon swapping interaction between the two resonators. By including the the direct resonator-resonator interaction, the net linear interaction can be switched off at a particular detuning \cite{Mariantoni08}. When an ideal qubit coupler is replaced by a transmon device, which is characterized by three (a qutrit) or more levels (a qudit), the form of the various interaction terms found above remains the same. However, the strength of the coupling coefficients now depends on the coupler anharmonicity~$\alpha = \omega_{01} - \omega_{12}$, where $\omega_{01}$ ($\omega_{12}$) is the level spacing in angular frequency between the ground state~$0$ and the first excited state~$1$ (the first and second excited state $1$ and $2$) of the transmon. In this case, we find that the nonlinear coupling strengths should be replaced by~$g^4 / \Delta^3 \to \chi / 4$, where
\begin{equation}
\chi \simeq - 2 \alpha \frac{g^4}{\Delta^4}
	\label{pert1}
\end{equation}
is found by perturbation theory~\cite{Footnote}.  Additional perturbative results can be found in the study of Ref.~\cite{Elliott2017}, where it is also shown that the nonlinear coupling coefficients can be switched off by introducing a second qubit in the coupling circuit.
\begin{figure}
\begin{center}
\includegraphics[width=3in]{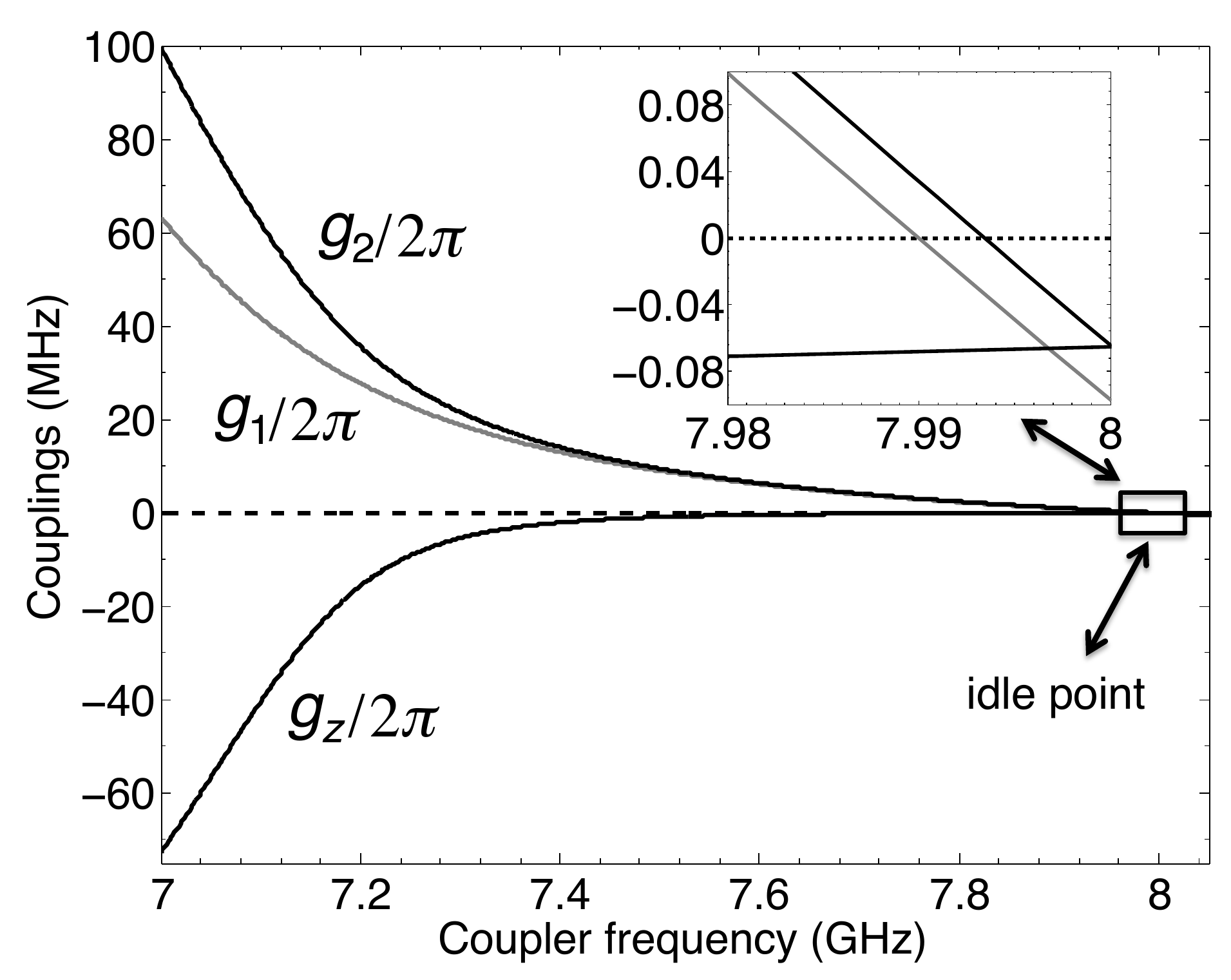}
\caption{Effective couplings~$g_1/2\pi$, $g_2/2\pi$, and $g_{z}/2\pi$ as a function of the (qutrit) coupler frequency $\omega_{01} / 2 \pi$ (see main text).}
	\label{couplings}
\end{center}
\end{figure}

\begin{figure*}[ht]
\begin{center}
\includegraphics[width=7in]{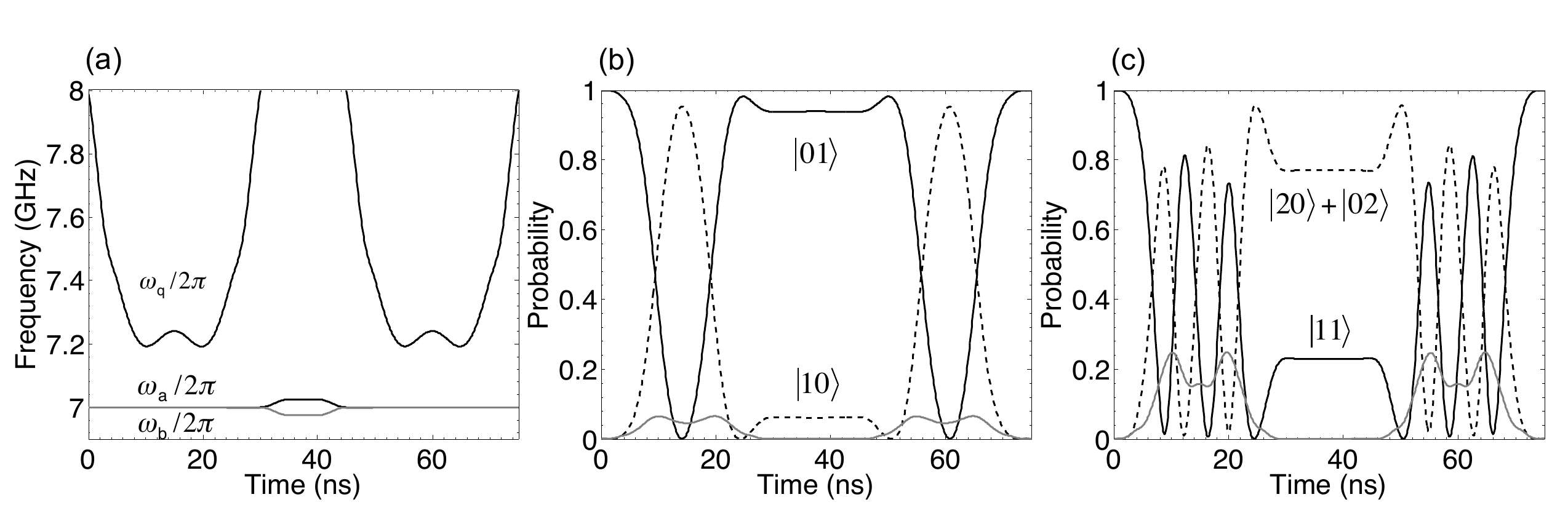}
\caption{Qubit logic gates. (a) The time-dependent shift of the coupler frequency~$\omega_{01} / 2 \pi$ (upper curve) and the resonators~$\omega_{\textrm{a}} / 2 \pi$ and $\omega_{\textrm{b}} / 2 \pi$ (lower curves) as a function of time. These control pulses are optimized to implement a \textsc{controlled-phase} gate with total time of~\SI{75}{\nano\second}. (b) State probabilities for the single-excitation subspace as a function of time, calculated using the time-dependent Schr{\"o}dinger equation with optimized control pulses and the initial two-resonator state~$\ket{01}$. The probabilities are for the two-resonator states~$\ket{01}$ (solid-black), $\ket{10}$ (dashed-black), and the excited state of the coupler (lower solid-gray curve). (c) State probabilities for the double-excitation subspace as a function of time, calculated using the optimized control pulses and the initial two-resonator state~$\ket{11}$. The probabilities are for the two-resonator states~$\ket{11}$ (solid-black), $\ket{20} + \ket{02}$ (dashed-gray), and excited states of the coupler
(lower curves) \cite{Footnote}.}
	\label{swap_fig}
\end{center}
\end{figure*}

Restricting our analysis to one or two photons, we numerically calculate the effective coupling coefficients~$g_1 \simeq  g^2 / \Delta - g_{\textrm{ab}}$, for $n = 1$ swapping, $g_2 \simeq g_1- \chi/2$, for $n = 2$ swapping, and an effective two-qubit Ising-type coupling~$g_z \simeq \chi$, as shown in Fig.~\ref{couplings} \cite{Footnote}. For the data in the figure, we use the physical parameters~$\omega / 2 \pi = \SI{7}{\giga\hertz}$, $\alpha / 2 \pi = \SI{300}{\mega\hertz}$, $g / 2 \pi = \SI{100}{\mega\hertz}$, and $g_{\textrm{ab}} / 2 \pi = \SI{10}{\mega\hertz}$. The figure shows that when the qubit is tuned by approximately~\SI{1}{\giga\hertz} away from both oscillators, there is a location where both~$g_1$ and $g_2$ are close to zero, which constitutes the idle point of the coupler. At that location, the Ising-type coupling~$g_z / 2 \pi$ is less than $\approx \SI{0.1}{\mega\hertz}$, consistent with the perturbative result of Eq.~(\ref{pert1}). However, by tuning the qubit close to the oscillator frequencies, the various coupling coefficients can be made as large as~\SI{100}{\mega\hertz}. Thus, the RQR system is characterized by an on-to-off coupling ratio of~$\approx 1000$. This ratio can be increased by adding an extra qubit to the coupling circuit, or by adjusting the idle point to larger detunings.

\textit{Qubit Logic Gates.}\textemdash To implement quantum logic gates in the RQR system, we must carefully tune the coupler frequency in time. We first consider a two-qubit implementation, in which the two resonators can be in the ground state~$\ket{00}$, either of the singly-excited states~$\ket{01}$ or $\ket{10}$, or the doubly-excited state~$\ket{11}$. To isolate a \textsc{controlled-phase} gate, we engineer a three-step control sequence, as shown in Fig.~\ref{swap_fig}~(a). In this sequence, we first tune the coupler near resonance ($\Delta \approx 0$) by a ``fast adiabatic'' pulse~\cite{Martinis2014}, allowing the single- and double-excitation subspaces to evolve for some interval of time. Since~$g_1 \ne g_2$, photons in the two subspaces swap between the resonators at different rates. We then return the coupler frequency to its idle point and shift the resonator frequencies by using dispersive interactions with the control qubits for a short time interval.  This step shifts the phases of the two subspaces by a factor of~$\pi$, which serves to reverse the single-excitation swapping. Finally, we tune the coupler back near resonance for an additional interval of time. The net effect is an overall evolution of the state~$\ket{11}$. By controlling the amplitude, shape, and duration of the coupler and resonator frequency shifts, a \textsc{controlled-phase} gate can be implemented (up to single-qubit phases \cite{Footnote}).

We numerically optimize the control pulses for this gate with a total time of~\SI{75}{\nano\second}. The simulation results displayed in Fig.~\ref{swap_fig}~(b) and (c) show the time evolution of the state probabilities calculated from the time-dependent Schr{\"o}dinger equation with initial states~$\ket{01}$ and $\ket{11}$, respectively. A full density matrix simulation yields a process fidelity~$\mathcal{F} \approx 0.999$, where we include an energy relaxation time for the resonators of~\SI{100}{\micro\second} as well as energy relaxation and dephasing times for the coupler of~\SI{40}{\micro\second} and \SI{30}{\micro\second}, respectively.

\begin{figure}[b]
\begin{center}
\includegraphics[width=3in]{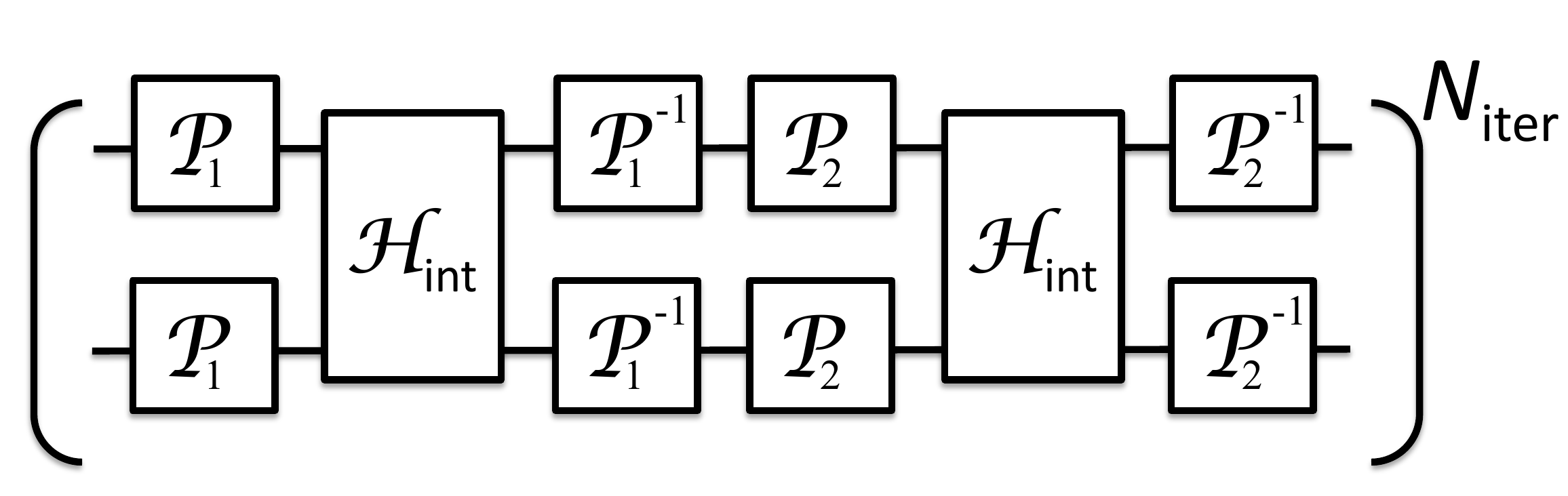}
\caption{Schematic quantum logic circuit to synthesize~$\mathcal{H}_{\subs{cross-Kerr}}$ from $\mathcal{H}_{\subs{int}}$.  For each iteration of this circuit, two interactions~$\mathcal{H}_{\subs{int}}$ are interleaved with qudit permutations~$\mathcal{P}$. This is repeated for some number of iterations~$N_{\subs{iter}}$, with permutations that depend on the iteration number (see main text).}
	\label{gate_sequence}
\end{center}
\end{figure}

\textit{Multiphoton Logic Gates.}\textemdash In the analysis above, we assume that the resonator-resonator detuning~$\delta = \omega_{\textrm{b}} - \omega_{\textrm{a}}$ is negligible. By including~$\delta$, however, we identify a new regime of operation. By tuning the coupler from large to small detuning, the state~$\ket{n_{\textrm{a}} = j , n_{\textrm{b}} = k}$ of modes~$R_{\textrm{a}}$ and $R_{\textrm{b}}$ can be mapped to the state~$\ket{n_- = j , n_+ = k}$ of the ``$+$'' and ``$-$'' modes, where~$j$ and $k$ are the possible photon numbers. The effective interaction between these modes can be found using second-perturbation theory for the Jaynes-Cummings states \cite{Footnote}.  For $\Delta = 0$ we find
\begin{equation}
\mathcal{H}_{\subs{int}} \simeq \frac{3}{4 \sqrt{2}} \frac{\delta^2}{g} n_- \sqrt{n_+}.
\label{ptint}
\end{equation}
While such an interaction does not directly produce the desired cross-Kerr interaction, the theory of Hamiltonian simulation ensures that it can be used to simulate it~\cite{Bennett2002}. We have designed a quantum logic circuit to synthesize $\mathcal{H}_{\subs{cross-Kerr}}$ from $\mathcal{H}_{\subs{int}}$ using set of single-qudit permutations, as shown in Fig.~\ref{gate_sequence}. The result of this sequence is an evolution according to the effective Hamiltonian
\begin{equation}
\mathcal{H}_{\subs{total}} = \sum_{j=1}^{N_{\subs{iter}}} \lambda_j \frac{1}{2} \left[\mathcal{P}_{j,1} \mathcal{H}_{\subs{int}} (\mathcal{P}_{j,1})^{-1} +  \mathcal{P}_{j,2} \mathcal{H}_{\subs{int}} (\mathcal{P}_{j,2})^{-1} \right],
\end{equation}
where~$\mathcal{P}_{j,k} = \mathcal{P}_{j,k}^{{\textrm{(a)}}} \otimes \mathcal{P}_{j,k}^{{\textrm{(b)}}}$ is a tensor product of permutations on each resonator, the weights~$\lambda_j$ are determined by the timing of the evolutions, and $N_{\subs{iter}}$ is the number of iterations of the circuit. We can choose the permutations, weights, and the number of iterations so that~$\mathcal{H}_{\subs{total}}$ approximates $\mathcal{H}_{\subs{cross-Kerr}}$ to high accuracy.

The single-resonator permutations can be implemented using each resonator's control qubit to realize single-resonator control protocols~\cite{Strauch2011, Strauch2012, Krastanov2015}. The simplest of these permutations is the transformation~$\ket{ j} \leftrightarrow \ket{N-j}$ on each resonator, where~$N$ is the maximum photon number of each resonator; the remaining permutations perform similar transformations on smaller number of photon states \cite{Footnote}.

Even for a single iteration of this sequence, with $\mathcal{P}_{1,1}$ the identity and $\mathcal{P}_{1,2}$ the transformation described above, a \textsc{controlled-phase} gate on states~$\ket{0}_{\textrm{L}}$ and $\ket{1}_{\textrm{L}}$ (with~$N = 4$) can be performed with high fidelity and with a short interaction time.  Using numerically calculated energy levels for the a qubit coupler with $g/2\pi = \SI{100}{\mega\hertz}$ and a resonator detuning of $\delta/2\pi = \SI{50}{\mega\hertz}$, we find that this gate can be operated at a variety of operation points with an ideal fidelity that is always greater than~$0.999$, and with total interactions times as short as~\SI{30}{\nano\second}. The same is true for a qutrit coupler, albeit with slightly longer interaction times.

\begin{figure}
\begin{center}
\includegraphics[width=3in]{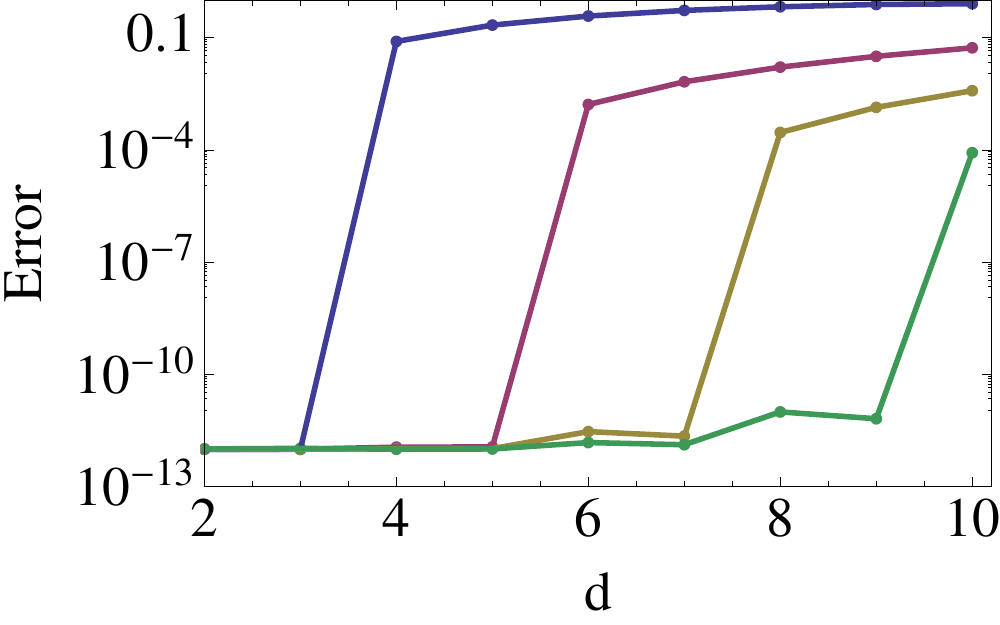}
\caption{(Color online) Gate error when implementing the multiphoton \textsc{controlled-phase} gate with~$\theta_d = 2 \pi /d$ for various dimensions~$d$ using the permutation circuit for a varying number of iterations~$N_{\subs{iter}}$.
The curves range from bottom to top with~$N_{\subs{iter}} = 1$ (blue), $N_{\subs{iter}} = 2$ (purple), $N_{\subs{iter}}=3$ (yellow), and $N_{\subs{iter}} = 4$ (green). Errors below $10^{-10}$ are effectively zero, subject only to numerical precision.}
	\label{cphase_error}
\end{center}
\end{figure}

As a final application of this quantum logic circuit, we observe that the cross-Kerr interaction can be used for logic gates on qudits. That is, turning on the interaction in Eq.~(\ref{eq1}) for a time~$t = \theta / \chi$ results in a multiply-\textsc{controlled-phase} gate
\begin{equation}
\ket{j} \ket{k} \to e^{i \theta j k} \ket{j} \ket{k} \,.
\end{equation}
Such gates, along with generalized Hadamard gates, can produce the multi-qudit quantum Fourier transform~\cite{Stroud2002} or, with~$\theta_d = 2 \pi / d$, the generalized \textsc{controlled-not} gate for qudit quantum computation~\cite{Gottesman99}. We calculate the process fidelity \cite{Footnote} of such gates using our permutation circuit to the interaction of Eq.~(\ref{ptint}) for the~$\theta_d$-gate for a variety of qudit dimensions~$d$ and iteration steps~$N_{\subs{iter}}$.  The resulting errors are shown in Fig.~\ref{cphase_error}. We see that each additional iteration increases the largest dimension of numerically exact gates by two, so that a $d$-dimensional gate requires approximately~$d$ applications of~$\mathcal{H}_{\subs{int}}$. This is more efficient than applying a sequence of two-state \textsc{controlled-phase} gates, one for each possible state~$\ket{j} \ket{k}$, which would require~$d^2$ interactions. Such a sequence was proposed in the theoretical work of Ref.~\cite{Strauch2011} and is similar to the approach taken in the recent experiment \cite{Rosenblum2018}.  Thus, this circuit produces  efficient high-fidelity entangling gates for resonator qudits.

\textit{Conclusion.}\textemdash In conclusion, we present a tunable coupling circuit that allows for nonlinear interactions between superconducting resonators. These interactions can be used to engineer high fidelity quantum logic operations on multiphoton two-resonator states.  For one such interaction, we devise a quantum circuit that, using single-resonator control, can be used to synthesize a strong cross-Kerr interaction for high-dimensional multiphoton states. We thus identify a promising route to full unitary control of multiple superconducting resonant cavities enabling a variety of applications to quantum information processing.

\acknowledgments
F.W.S. acknowledges the support of the Institute for Quantum Computing, where some of this work was performed. M.M. acknowledges the Natural Sciences and Engineering Research Council of Canada (NSERC) for financial support.

\beginsupplement
\newpage

\begin{widetext}

\section{Supplemental Material for ``Multiphoton Quantum Logic Gates for Superconducting Resonators with Tunable Nonlinear Interaction''}

\section{SI. Introduction}

In this supplemental material, we provide further details on the quantum-mechanical calculations presented in the main text. First, we diagonalize the Hamiltonian of the resonator-coupler-resonator (RCR) system shown in Fig.~1 of the main text both for a quantum bit (qubit) and a three-level system (qutrit) coupler in sections SII and SIII, respectively. We provide further details on the properties of the RCR system including analytical approximations and the numerical calculations leading to Fig.~2 of the main text. Second, in Sec. SIV, we provide further details of the analytical and numerical calculations used to study the qubit logic gate presented in and shown in Fig.~3 of the main text. Finally, in Sec. SV, we further elucidate the permutation circuit shown in Fig.~4 of the main text, and used to operate the multiphoton logic gates described by Fig. 5 of the main text.

\section{SII. RCR System: Qubit Coupling}

\subsection{A. Model Hamiltonian}

The Hamiltonian used to model a qubit coupler interacting with resonators~$R_{\textrm{a}}$ and $R_{\textrm{b}}$ reads (with~$\hbar = 1$):
\begin{equation}
\mathcal{H} = \omega_{\textrm{a}} a^{\dagger} a  + \omega_{\textrm{q}} \sigma^{\dagger} \sigma + \omega_{\textrm{b}} b^{\dagger} b + \ g_{\textrm{ab}} \left(a^{\dagger} b + b^{\dagger} a \right) + g_{\textrm{a}} \left(a^{\dagger} \sigma + a \sigma^{\dagger} \right) + g_{\textrm{b}} \left( b^{\dagger} \sigma + b \sigma^{\dagger} \right) \, .
\end{equation}
As described in the main text, the case~$\omega_{\textrm{a}} = \omega_{\textrm{b}} = \omega$, $g_{\textrm{a}} = g_{\textrm{b}} = g$, and $g_{\textrm{ab}} = 0$ can be solved by introducing the normal mode operators
\begin{equation}
c_{\pm} = \frac{1}{\sqrt{2}} \left( a \pm b \right)
\end{equation}
so that 
\begin{equation}
\mathcal{H}_0 = \omega \, c_{-}^{\dagger} c_{-} + \omega_{\textrm{q}} \, \sigma^{\dagger} \sigma + \omega \, c_{+}^{\dagger} c_{+} + \sqrt{2} g \left( c_{+}^{\dagger} \sigma + c_{+} \sigma^{\dagger} \right) \, .
	\label{Hzero}
\end{equation}
The eigenstates of~$\mathcal{H}_0$ are
\begin{subnumcases}{}
|\Psi_{n_-, n_+}^{(+)}\rangle = |n_-\rangle \left( \sin \theta_{n_+}  |n_+, 0\rangle +\cos \theta_{n_+} |n_+-1,1\rangle \right) \, , \\
|\Psi_{n_-, n_+}^{(-)}\rangle = |n_-\rangle \left(\cos \theta_{n_+} |n_+, 0\rangle - \sin \theta_{n_+}  |n_+-1,1\rangle \right) \, ,
\end{subnumcases}
with eigenvalues
\begin{equation} 
E_{n_-, n_+, \pm} = \left(n_- + n_+\right) \omega + \frac{1}{2} \left( \Delta \pm \sqrt{ \Delta^2 + 8 n_+ g^2} \right) \, ,
\end{equation}
where~$\tan (2 \theta_{n_+}) = 2 \sqrt{2 n_+} g / \Delta$, $\Delta = \omega_{\textrm{q}} - \omega$ and we are using a basis~$|n_-, n_+, n_{\textrm{q}} \rangle$ for the states with~$n_-$ excitations in mode ``$-$'', $n_+$ excitations in mode ``$+$'', and $n_{\textrm{q}}$ excitations in the qubit. For~$\Delta > 0$, the state amplitudes can be expressed as
\begin{subnumcases}{}
\sin \theta_{n_+} = \frac{1}{\sqrt{2}} \left(1 - \frac{\Delta}{\sqrt{\Delta^2 + 8 n_+ g^2}} \right)^{1/2} \, , \\
\cos \theta_{n_+} = \frac{1}{\sqrt{2}} \left(1 + \frac{\Delta}{\sqrt{\Delta^2 + 8 n_+ g^2}} \right)^{1/2} \, .
\end{subnumcases}
By working in the dressed basis with the coupler in its ground state, the effective Hamiltonian can be obtained by replacing~$n_+ \to c_+^{\dagger} c_+$ and $n_- \to c_-^{\dagger} c_-$ in $E_{n_-, n_+,-}$, leading to Eq.~(5) in the main text.

\subsection{B. Effect of Direct Resonator-Resonator Interaction and Detuning}

In the model Hamiltonian~$\mathcal{H}_0$ of Eq.~(\ref{Hzero}), we set $g_{\textrm{ab}} = 0$ and $\omega_{\textrm{a}} = \omega_{\textrm{b}} = \omega$. If we relax these assumptions, the Hamiltonian can be written as
\begin{equation}
\mathcal{H} = \mathcal{H}_0 + \mathcal{H}_1 + \mathcal{H}_2 \, ,
\end{equation}
where~$\omega = (\omega_{\textrm{a}} + \omega_{\textrm{b}}) / 2$ in $\mathcal{H}_0$, 
\begin{equation}
\mathcal{H}_1 =  g_{\textrm{ab}} \left( a^{\dagger} b + b^{\dagger} a \right)
\end{equation}
is the direct resonator-resonator interaction, and
\begin{equation}
\mathcal{H}_2 = \frac{1}{2} \delta \left( b^{\dagger} b - a^{\dagger} a \right)
\end{equation}
is due to the detuning~$\delta = \omega_{\textrm{b}} - \omega_{\textrm{a}}$ of the resonators. Including these terms leads to additional terms in the effective Hamiltonian described by Eqs.~(5)-(7) in the main text.

We first consider the direct resonator-resonator interaction term~$\mathcal{H}_1$. This can be rewritten in terms of the ``$+$'' and ``$-$'' modes as
\begin{equation}
\mathcal{H}_1 = g_{ab} \left( c_{+}^{\dagger} c_+ - c_-^{\dagger} c_-\right) \, .
\end{equation}
Using first-order perturbation theory for the states~$|\Psi_{n_-, n_+}^{(-)}\rangle$, we find
\begin{equation}
\Delta E_1 = \langle \Psi_{n_-, n_+}^{(-)} | \mathcal{H}_1 | \Psi_{n_-, n_+}^{(-)} \rangle =  g_{ab} \left(n_+ - \sin^2 \theta_{n_+} - n_-\right) \, .
\end{equation}
Using the dressed basis, we again make the replacements~$n_+ \to c_+^{\dagger} c_+$ and $n_- \to c_-^{\dagger} c_-$ so that the effective Hamiltonian corresponding to~$\mathcal{H}_1$ becomes
\begin{eqnarray}
\mathcal{H}_{1, \subs{eff}} & = & g_{\textrm{ab}} (c_{+}^{\dagger} c_{+} - c_{-}^{\dagger}c_-) - \frac{1}{2} g_{\textrm{ab}} \left( 1 - \frac{\Delta}{\sqrt{\Delta^2 + 8 g^2 c_+^{\dagger} c_+}} \right)  \nonumber \\
& \simeq & g_{ab} (c_+^{\dagger} c_+ - c_-^{\dagger} c_-) + 2 g_{\textrm{ab}} \frac{g^2}{\Delta^2} c_+^{\dagger} c_+ - 12 g_{\textrm{ab}} \frac{g^4}{\Delta^4} (c_+^{\dagger} c_+)^2.
	\label{ptdirect}
\end{eqnarray}
For the RCR system considered in this work, we typically have~$g_{\textrm{ab}} \ll g \ll \Delta$. Thus, the terms proportional to $g^2/\Delta^2$ and $g^4/\Delta^4$ in~$\mathcal{H}_{1, \subs{eff}}$ can be safely ignored.

We now consider the resonator-resonator detuning term~$\mathcal{H}_2$.  This can be rewritten in terms of the ``$+$'' and ``$-$'' modes as
\begin{equation}
\mathcal{H}_2 = - \frac{\delta}{2} \left( c_+^{\dagger} c_- + c_-^{\dagger} c_+ \right) \, .
\end{equation}
Assuming that~$\delta < g$, this can be treated using second-order perturbation theory.  $\mathcal{H}_2$ couples a given state~$|\Psi_{n_-, n_+}^{(-)}\rangle$ to the four states~$|\Psi_{n_- -1, n_+ +1}^{(-)}\rangle$, $|\Psi_{n_- + 1, n_+ - 1}^{(-)}\rangle$, $|\Psi_{n_- - 1, n_+ + 1}^{(+)}\rangle$, $|\Psi_{n_- + 1, n_+ -1}^{(+)}\rangle$. The dominant contribution comes from the first two states, for which the perturbative shift is given by
\begin{equation}
\Delta E_2 = \frac{1}{2} \delta^2 \left[ \frac{n_- \left( A_{n_+}  \sqrt{n_+ +1} + B_{n_+}  \sqrt{n_+}\right)^2}{\sqrt{\Delta^2 + 8 (n_+ + 1) g^2} - \sqrt{\Delta^2 + 8 n_+ g^2}} + \frac{(n_- + 1) \left( A_{n_+-1} \sqrt{n_+} + B_{n_+-1} \sqrt{n_+-1}\right)^2}{\sqrt{\Delta^2 + 8 (n_+ - 1) g^2} - \sqrt{\Delta^2 + 8 n_+ g^2}}\right] \, ,
	\label{ptshift}
\end{equation}
where~$A_n = \cos \theta_n \cos \theta_{n+1}$, $B_n = \sin \theta_n \sin \theta_{n+1}$, and $\Delta = \omega_{\textrm{q}} - (\omega_{\textrm{a}} + \omega_{\textrm{b}}) / 2$.

There are two natural limits for this energy shift. For~$\Delta \to 0$, we have~$A_n= B_n = 1/2$ and thus
\begin{eqnarray}
\left( \Delta E_2 \right)_{\Delta \to 0} & = & \frac{1}{16 \sqrt{2}} \frac{\delta^2}{g} \left [ \frac{ n_- \left(\sqrt{n_+ +1} + \sqrt{n_+}\right)^2}{\sqrt{n_+ + 1} - \sqrt{ n_+ }} + \frac{(n_- + 1) \left( \sqrt{n_+} +  \sqrt{n_+-1}\right)^2}{\sqrt{n_+ - 1} - \sqrt{n_+}}\right] \nonumber \\
 & = & \frac{1}{16 \sqrt{2}} \frac{\delta^2}{g} \left[ n_- \left(\sqrt{n_+ +1} + \sqrt{n_+}\right)^3 - (n_- + 1) \left( \sqrt{n_+} +  \sqrt{n_+-1}\right)^3 \right] \, .
\end{eqnarray}
It can be shown that
\begin{equation}
\left( \sqrt{n_+ +1} + \sqrt{n_+}\right)^3 - \left( \sqrt{n_+} +  \sqrt{n_+-1}\right)^3 \simeq 12 \sqrt{n}
\end{equation}
to high accuracy. Using this expression the energy shift can be approximated by
\begin{equation}
\left( \Delta E_2 \right)_{\Delta \to 0} \simeq \frac{1}{16 \sqrt{2}} \frac{\delta^2}{g} \left[ 12 n_- \sqrt{n_+} - \left(\sqrt{n_+} +  \sqrt{n_+-1}\right)^3 \right] \, .
	\label{hint}
\end{equation}
The resonator-resonator interaction described in the main text arises from the first term in Eq.~(\ref{hint}), while the second term provides a single-resonator phase shift. For~$\Delta \gg g$, a Taylor series expansion of Eq.~(\ref{ptshift}) in powers of~$g/\Delta$, using~$A_n \simeq 1 - (2 n + 1) g^2 / \Delta^2$ and $B_n \simeq 2 \sqrt{n (n+1)} g^2 / \Delta^2$, results in 
\begin{equation}
\left( \Delta E_2 \right)_{\Delta \gg g} \simeq \frac{\delta^2}{16 g} \left( 2 \frac{\Delta}{g} n_- + 16 \frac{g}{\Delta} n_- n_+ \right) \, .
	\label{hint2}
\end{equation}
In this limit, there is a {\em bona fide} cross-Kerr interaction from the second term in Eq.~(\ref{hint2}), albeit with a coupling coefficient of~$\delta^2 / \Delta \ll \delta^2 / g$.

\section{SIII. RCR System: Qutrit Coupler}

\subsection{A. Model Hamiltonian}
A qutrit coupler can be realized by means of a superconducting transmon device, which is characterized by more than two quantized levels~\cite{Koch07s}. The Hamiltonian used to model a qutrit coupler interacting with resonators~$R_{\textrm{a}}$ and $R_{\textrm{b}}$ reads (with $\hbar = 1$):
\begin{equation}
\mathcal{H} = \omega_{\textrm{a}} a^{\dagger}a + \mathcal{H}_{\textrm{q}} + \omega_{\textrm{b}} b^{\dagger} b + \ g_{\textrm{ab}} \left(a^{\dagger} b + b^{\dagger} a \right) + g_{\textrm{a}} \left(a^{\dagger} \sigma + a \sigma^{\dagger} \right) + g_{\textrm{b}} \left( b^{\dagger} \sigma + b \sigma^{\dagger} \right) \, ,
\end{equation}
where~$\sigma = |0\rangle \langle 1| + \sqrt{2} |1\rangle \langle 2|$,
\begin{equation}
\mathcal{H}_{\textrm{q}} = \left( \begin{array}{ccc}
0 & 0 & 0 \\
0 & \omega_{01} & 0 \\
0 & 0 & \omega_{01} + \omega_{12} 
\end{array} \right) = \left( \begin{array}{ccc} 
0 & 0 & 0 \\
0 & \omega_{01} & 0 \\
0 & 0 & 2 \omega_{01} - \alpha \end{array} \right) \, ,
\end{equation}  
$\alpha = \omega_{01} - \omega_{12}$ is a fixed anharmonicity, $\omega_{01}$ is the tunable angular frequency for the transition from the ground state~$|0\rangle$ to the first excited state~$|1\rangle$ of the qutrit, and $\omega_{12}$ is the angular frequency for the transition from the first excited state $|1\rangle$ to the second excited state~$|2\rangle$.

For the case~$\omega_{\textrm{a}} = \omega_{\textrm{b}} = \omega$, $g_{\textrm{a}} = g_{\textrm{b}} = g$, and $g_{\textrm{ab}} = 0$, and using the modes $c_{\pm}$, we find
\begin{equation}
\mathcal{H} = \omega c_{-}^{\dagger} c_{-} + \mathcal{H}_{\textrm{q}} + \omega c_{+}^{\dagger} c_{+} + \sqrt{2} g \left(c_{+}^{\dagger} \sigma + c_+ \sigma \right) \, .
\end{equation}
In principle, we can solve for the eigenvalues and eigenvectors analytically but, in practice, it is more convenient to perform perturbative or numerical calculations.

\subsection{B. Pertubative analysis}
When~$\Delta \gg g$, we can calculate the perturbative energy shift of the state~$|n_-, n_+, n_{\textrm{q}}\rangle$ by considering the reduced Hamiltonian
\begin{equation}
 \left( \begin{array}{ccc}
(n_+ + n_-) \omega & \sqrt{2 n_+} g & 0 \\
\sqrt{2 n_+} g & (n_+ + n_-) \omega + \Delta & 2 \sqrt{n_+-1} g \\
0 & 2 \sqrt{n_+-1} g & (n_+ + n_-) \omega + 2 \Delta - \alpha 
\end{array} \right) \, ,
\end{equation}
where we use the basis states~$|n_-, n_+, 0\rangle$, $|n_-, n_+-1,1\rangle$ and $|n_-, n_+-2,2\rangle$. By means of fourth-order perturbation theory in $g$, we find that
\begin{equation}
E_{n_-, n_+, n_{\textrm{q}} =0} \simeq  (n_- + n_+) \omega - n_+ \left( \frac{2 g^2}{\Delta} - \frac{4 g^4}{ \Delta^2 (\Delta - \alpha/2)} \right) + n_+^2 \left( \frac{4 g^4}{\Delta^3} - \frac{4 g^4}{\Delta^2 (\Delta - \alpha/2)} \right) \, .
	\label{pteq1}
\end{equation}
The last term in Eq.~(\ref{pteq1}) is the Kerr shift, which can be simplified when~$\Delta \gg \alpha$,
\begin{equation}
\chi = \left( \frac{4 g^4}{\Delta^3} - \frac{4 g^4}{\Delta^2 (\Delta - \alpha/2)} \right) \approx - 2 \alpha \frac{g^4}{\Delta^4} \, .  
\end{equation}

\subsection{C. Numerical analysis}
For identical resonators with~$\omega_{\textrm{a}} / 2\pi = \omega_{\textrm{b}} / 2\pi = \SI{7}{\giga\hertz}$, couplings~$g_{\textrm{a}} / 2\pi = g_{\textrm{b}} / 2\pi = \SI{100}{\mega\hertz}$, $g_{\textrm{ab}} / 2\pi = \SI{10}{\mega\hertz}$, and $\alpha / 2\pi = \SI{300}{\mega\hertz}$, we numerically  calculate the eigenstates~$|v_n\rangle$ and eigenvalues~$E_n$ of the system, where~$n = 0, 1, 2, \dots \ $. The energy levels are shown in Fig.~\ref{elevel1}, where we also indicate the composition of the various eigenstates by~$|n_{\textrm{a}}, n_{\textrm{b}}, n_{\textrm{q}}\rangle$ corresponding to those states with~$n_{\textrm{a}}$ excitations in~$R_{\textrm{a}}$, $n_{\textrm{b}}$ excitations in~$R_{\textrm{b}}$, and $n_{\subs{q}}$ excitations in the coupler. The eigenstates for large~$\omega_{01}$ are approximately given by
\begin{eqnarray}
|v_0\rangle & = & |000\rangle \nonumber \\
|v_1\rangle & = & \frac{1}{\sqrt{2}}  \left( |100\rangle + |010\rangle \right) \nonumber \\
|v_2\rangle & = & \frac{1}{\sqrt{2}} \left( |100\rangle - |010\rangle \right)  \nonumber \\
|v_3\rangle & = &  |001\rangle \nonumber \\
|v_4\rangle & = & \frac{1}{2}  \left( |200\rangle + |020\rangle + \sqrt{2} |110 \rangle \right)  \nonumber \\ 
|v_5\rangle & = & \frac{1}{\sqrt{2}} \left( |200\rangle - |020\rangle \right) \nonumber \\
|v_6\rangle & = & \frac{1}{2} \left( |200\rangle + |020\rangle - \sqrt{2} |110\rangle \right) \nonumber \\
|v_7\rangle & = & \frac{1}{\sqrt{2}} \left( |101\rangle + |011\rangle \right) \nonumber \\
|v_8\rangle & = & \frac{1}{\sqrt{2}} \left( |101\rangle - |011\rangle \right) \nonumber \\
|v_{9}\rangle & = & |002\rangle \, \nonumber .
\end{eqnarray} 
The natural two-qubit states are given by the superpositions
\begin{equation}
\begin{cases}
|00\rangle_{\subs{qubit}} = |v_0\rangle \, , \\
|01\rangle_{\subs{qubit}} = \dfrac{1}{\sqrt{2}} (|v_1\rangle - |v_2\rangle) \, , \\
|10\rangle_{\subs{qubit}} = \dfrac{1}{\sqrt{2}} (|v_1\rangle + |v_2\rangle) \, ,  \\
|11\rangle_{\subs{qubit}} = \dfrac{1}{\sqrt{2}} (|v_4\rangle - |v_6\rangle) \, ,
\end{cases}
	\label{qubitbasis}
\end{equation}
where the coupler is in its ground state. Note that for finite detuning, the interaction with the coupler leads to a dressing of these states by terms proportional to~$g/\Delta$.

\begin{figure}
\includegraphics[width=6in]{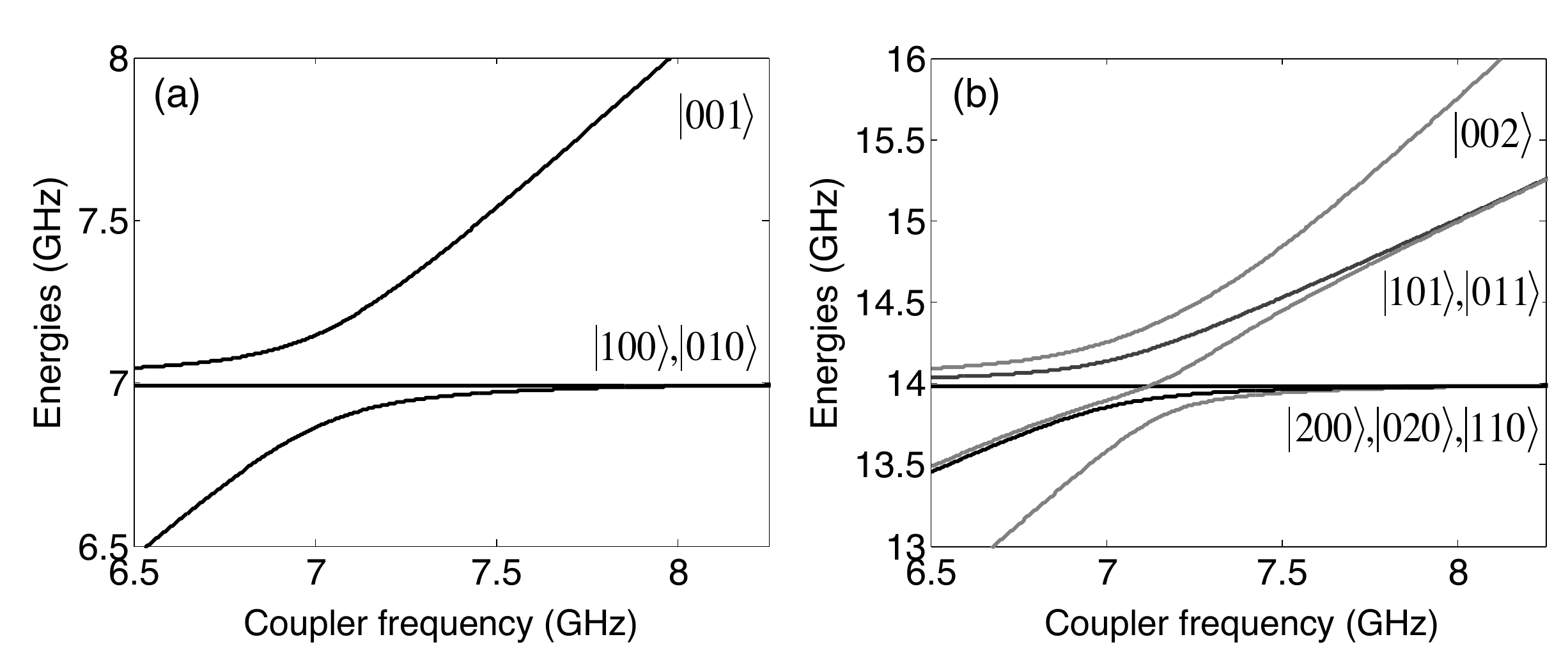}
\caption{Energy eigenvalues~$E_n / 2\pi$ of the (a) first excited states ($n = 1\mbox{--}3$) and (b) second excited states ($n = 4\mbox{--}9$) of the RCR system as a function of the qutrit coupler frequency~$\omega_{01} / 2\pi$. The state labels~$|n_{\textrm{a}}, n_{\textrm{b}}, n_{\textrm{q}}\rangle$ indicate the state compositions for large values of~$\omega_{01}/2\pi$ (see main text).}
	\label{elevel1}
\end{figure}

We define the coupling coefficients for the two-qubit states of Eq.~(\ref{qubitbasis}) by the energy splittings, as follows:
\begin{equation}
\begin{cases}
g_1 = (E_2 - E_1) / 2 \, , \\
g_2 = (E_6 - E_4) / 4 \, , \\
g_{z} = \left( \dfrac{1}{2} E_6 + \dfrac{1}{2} E_4 + E_0 - E_1 - E_2 \right) \, .
\end{cases}
\end{equation}
The splitting of~$2 g_1$ between states~$|v_1\rangle$ and $|v_2\rangle$ corresponds to the linear coupling between resonators~$R_{\textrm{a}}$ and $R_{\textrm{b}}$. The splitting of~$4 g_2$ between states~$|v_4\rangle$ and $|v_6\rangle$ differs from that due to a linear coupler~($4 g_1$) because of the number-state-dependent swapping term.  Finally, the Ising-type two-qubit coupling~$g_{z}$ arises from the Kerr shift. The numerically calculated values of these coupling coefficients are shown in Fig.~2 of the main text.

We can also approximate these values by using the perturbative results of Eqs.~(\ref{ptdirect}) and (\ref{pteq1}). We find
\begin{equation}
\begin{cases}
E_0 = E_{n_- = 0, n_+ = 0, n_{\textrm{q}} = 0} = 0 \, , \\
E_1 \simeq E_{n_- = 0, n_+ = 1, n_{\textrm{q}} = 0} + g_{\textrm{ab}} \simeq \omega - \dfrac{2 g^2}{\Delta} + \dfrac{4 g^4} {\Delta^3} + g_{\textrm{ab}} \, , \\
E_2 \simeq E_{n_- = 1, n_+ = 0, n_{\textrm{q}} = 0} - g_{\textrm{ab}} \simeq \omega - g_{\textrm{ab}} \, , \\
E_4 \simeq E_{n_- = 0, n_+ = 2, n_{\textrm{q}} = 0} + 2 g_{\textrm{ab}} \simeq 2 \omega - \dfrac{4 g^2}{\Delta} + \dfrac{8 g^4}{\Delta^3} - 4 \alpha \dfrac{g^4}{\Delta^4} + 2 g_{\textrm{ab}} \, , \\
E_6 \simeq E_{n_- = 2, n_+ = 0, n_{\textrm{q}} = 0} - 2 g_{\textrm{ab}} \simeq 2 \omega - 2 g_{\textrm{ab}} \, ,
\end{cases}
\end{equation}
so that
\begin{equation}
\begin{cases}
g_1 \simeq \dfrac{g^2}{\Delta} - g_{\textrm{ab}} - \dfrac{2g^4}{\Delta^3} \, , \\
g_2 \simeq \dfrac{g^2}{\Delta} - g_{\textrm{ab}} - \dfrac{2 g^4}{\Delta^3} + \alpha \dfrac{g^4}{\Delta^4} \, , \\
g_z \simeq - 2 \alpha \dfrac{g^4}{\Delta^4} \, .
\end{cases}
\end{equation}
Thus, we have $g_2 \simeq g_1 - \chi/2$ and $g_z \simeq \chi$. 

\section{SIV. Two-Qubit Logic Gates}

By turning on the coupling, a large variety of entangling evolutions can be generated. These have the generic form
\begin{equation}
U(\theta,\phi) = \left( \begin{array}{cccc}
1 & 0 & 0 & 0 \\
0 & \cos \theta & -i \sin \theta & 0 \\
0 & -i \sin \theta & \cos \theta & 0 \\
0 & 0 & 0 & e^{-i \phi}
\end{array} \right) \, ,
\end{equation}
where~$\theta$ is a swap angle between states~$|01\rangle$ and $|10\rangle$, $\phi$ is a controlled phase on~$|11\rangle$, and we have removed any single-qubit phases~\cite{Strauch2003}. To isolate a \textsc{controlled-phase} gate, we can interrupt the entangling evolution by means of a one-qubit phase gate~$Z_1$ and the following identity:
\begin{equation}
U(0,2\phi) = Z_1 U(\theta,\phi) Z_1 U(\theta,\phi) \, .
	\label{gateidentity}
\end{equation}
If $\phi = \pi/2$ we have $U(0,\pi) = U_{\textrm{CZ}}$. This approach leads us to engineer a three-step control sequence for the frequencies of the qutrit coupler and the two resonators.

We choose two control pulses ~$\Delta \omega_{\textrm{q}}(t)$ and $\Delta \omega_{\textrm{ab}}(t)$ that shift the coupler frequency in time by~$\omega_{01}(t) = \omega_{01,\subs{idle}}  + \Delta \omega_{\textrm{q}}(t)$ and the two resonator frequencies by~$\omega_{\textrm{a}}(t) = \omega + \Delta \omega_{\textrm{ab}}(t)$ and $\omega_{\textrm{b}}(t) = \omega - \Delta \omega_{\textrm{ab}}(t)$. The resonator frequencies are controlled by dispersive interactions with the corresonding control qubits, which are indicated in Fig.~1 of the main text. The control pulses are defined by
\begin{equation}
\Delta \omega_{\textrm{q}} (t) = \Delta \omega_1 \left\{ \begin{array}{lll}  t / \tau_1 +  \sum_{j=1}^3 a_j \sin(j \pi t/\tau_1) & \mbox{for} & 0< t < \tau_1 \\
1 + \sum_{j=1}^3 b_j \sin[j \pi (t-\tau_1) / T_1] & \mbox{for} & \tau_1 < t < T _1+ \tau_1 \\
\tilde{t} / \tau_1 + \sum_{j=1}^3 a_j \sin(j \pi \tilde{t}_1/\tau) & \mbox{for} & T_1 + \tau_1 < t < T_1 + 2\tau_1 \\
\end{array} \right .
\end{equation}
where~$\tilde{t}_1 = T_1 + 2 \tau_1 - t$ and
\begin{equation}
\Delta \omega_{\textrm{ab}} (t) = \Delta \omega_2 \left\{ \begin{array}{lll} \dfrac{1}{2} \left[ 1 - \cos (\pi t / \tau_2 ) \right] & \mbox{for} & 0< t < \tau_2 \\
1 & \mbox{for} & \tau_2 < t < T_2 + \tau_2 \\
 \dfrac{1}{2} \left[ 1 - \cos (\pi \tilde{t}_2 / \tau \right] & \mbox{for} & T_2 + \tau_2 < t < T_2 + 2\tau_2 \\
\end{array} \right  .
\end{equation}
where $\tilde{t}_2 = T_2+ 2 \tau_2 - t$. To optimize these controls, we choose values for~$T_1, \tau_1, T_2, \mbox{and} \ \tau_2$ and numerically search for the coefficients~$\{ a_j \}$ and $\{ b_j \}$ with the objective of finding a gate equivalent to~$U_{\subs{CZ}}$, up to single-qubit phases (which can be added to the ends of the overall sequence). The numerically optimized pulse parameters are found to be~$\Delta \omega_1 / 2\pi = \SI{-0.7153}{\giga\hertz}$, $\tau_1 = \SI{5}{\nano\second}$,  $T_1 = \SI{20}{\nano\second}$, $\{a_j\} = \{0.1637, -0.0974, -0.0372\}$, $\{b_j\} = \{0.1017, 0.0078, 0.0131\}$, $\Delta \omega_2 / 2\pi = \SI{0.0252}{\giga\hertz}$,  $\tau_2 = \SI{5}{\nano\second}$,  and $T_2 = \SI{5}{\nano\second}$. This gate has a total time~$T_{\subs{gate}} = 2 (T_1 + 2 \tau_1) + T_2 + 2 \tau_2 = \SI{75}{\nano\second}$ and has process fidelity of approximately~$0.99999$. These pulses are shown in Fig.~3(a) in the main text.

Using these control pulses, we numerically simulate the time-dependent Schr{\"o}dinger equation to find~$|\Psi(t)\rangle$ with starting with various initial conditions chosen from the qubit states given by Eq.~(\ref{qubitbasis}) using the eigenstates calculated at the idle point of the coupler. First, we set~$|\Psi(t=0)\rangle = |01\rangle_{\subs{qubit}}$ and calculate the probabilities for the qubit states~$|\langle \Psi(t)| 01\rangle_{\subs{qubit}}|^2$ and $|\langle \Psi(t)| 10\rangle_{\subs{qubit}}|^2$, as well as the probability of excitation in the coupler~$P_{\subs{excite}} = |\langle \Psi(t) | v_3\rangle|^2$; these are shown in Fig.~3(b) of the main text. The dominant oscillation is the swap~$|010\rangle \to |100\rangle$, which is reversed by the intermediate~$Z$-gate. There is also a small coupling to state~$|v_3\rangle = |001\rangle$, in which the coupler is excited, but no residual population at~$t=T_{\subs{gate}}$. Second, we set~$|\Psi(t=0)\rangle = |11\rangle_{\subs{qubit}}$ and calculate~$|\langle \Psi(t)| 11\rangle_{\subs{qubit}}|^2$, $|\langle \Psi(t) | \bar{v}\rangle|^2$, where $|\bar{v}\rangle = (|200\rangle + |020\rangle)/\sqrt{2}$, as well as $P_{\subs{excite}} = |\langle \Psi(t) | v_7\rangle|^2 +|\langle \Psi(t)| v_9 \rangle|^2 $. These three probabilities are shown in Fig.~3(c) of the main text. Here, the dominant oscillation is between~$|110\rangle$ and $(|200\rangle + |020\rangle) / \sqrt{2}$. There is also small couplings to the states~$|v_7\rangle$ and $|v_9\rangle$ but, again, no residual population at~$t = T_{\subs{gate}}$. 
We also simulate the full gate dynamics using the Lindblad equation for the density matrix~$\rho$,
\begin{equation}
\frac{ d\rho}{dt} = -i [\mathcal{H},\rho] + \sum_j \lambda_j \left(L_j \rho L_j^{\dagger} - \frac{1}{2} L_j^{\dagger} L_j \rho - \frac{1}{2} \rho L_j^{\dagger} L_j \right) \, ,
\end{equation}
using four Lindblad operators
\begin{equation}
\begin{cases}
\lambda_1 = 1/T_r \ \mbox{and} \ L_1 = a \, , \\
\lambda_2 = 1/T_r \ \mbox{and} \ L_2 = b \, \\
\lambda_3 = 1/T_q \ \mbox{and} \ L_3 = \sigma \, , \\
\lambda_4 = 2/T_{\varphi} \ \mbox{and} \ L_4 = \sigma^{\dagger} \sigma \, ,
\end{cases}
\end{equation}
with~$T_r = \SI{100}{\micro\second}$, $T_q = \SI{40}{\micro\second}$, $T_{\varphi} = \SI{30}{\micro\second}$. The first three Lindblad operators correspond to energy relaxation of resonator~$R_{\textrm{a}}$, resonator $R_{\textrm{b}}$, and the coupler, respectively, while the last corresponds to dephasing of the coupler. By projecting the density matrix onto the two-qubit states (defined as above), the full process matrix is extracted with process fidelity~$0.9987$ and leakage~$4 \times 10^{-4}$.

\section{SV. Multiphoton Controlled-Phase Gates}

\subsection{A. Permutation Circuit}

Here, we specify how to synthesize a cross-Kerr interaction from the interaction
\begin{equation}
\mathcal{H}_{\subs{int}} = \chi n_- (n_+)^{1/2}
\end{equation}
using the permutation circuit shown in Fig. 4 of the main text. The basic principle is to split the evolution into a sequence of iterations in which the qudit states (treating the resonators as $d$-dimensional systems with computational states $|0\rangle, \dots, |d-1\rangle$) are permuted and the interactions timed so as to ``linearize'' the total interaction with respect to $n_+$. For each iteration, the effective Hamiltonian becomes
\begin{equation}
\mathcal{H}_j = \frac{1}{2} \mathcal{P}_{j,1} \mathcal{H}_{\subs{int}} (\mathcal{P}_{j,1})^{-1} +  \frac{1}{2} \mathcal{P}_{j,2} \mathcal{H}_{\subs{int}} (\mathcal{P}_{j,2})^{-1} \, ,
\end{equation}
where~$\mathcal{P}_{j,k} = \mathcal{P}_{j,k}^{{\textrm{(a)}}} \otimes \mathcal{P}_{j,k}^{{\textrm{(b)}}}$ is a tensor product of permutations on each resonator. The total effect of the circuit is to apply a linear combination of the Hamiltonians
\begin{equation}
\mathcal{H}_{\subs{total}} = \sum_{j=1}^{N_{\subs{iter}}} \lambda_j \mathcal{H}_j \, ,
\end{equation}
with permutations~$\mathcal{P}_{j,1}, \mathcal{P}_{j,2}$ and weights~$\lambda_j$ chosen to approximate the cross-Kerr interaction as closely as possible.

The first iteration uses the permutations~$\mathcal{P}_{j=1,1} = \mathcal{I} \otimes \mathcal{I}$ and
$\mathcal{P}_{j=1,2} = \mathcal{P}_1 \otimes \mathcal{P}_1$, where~$\mathcal{I}$ is the identity and $\mathcal{P}_1$ flips the photon numbers by
\begin{equation}
\mathcal{P}_1 |j\rangle = |N-j\rangle \, ,
\end{equation}
where~$N$ is the maximum photon number under consideration. The resulting effective Hamiltonian is
\begin{equation}
\mathcal{H}_{j=1} = \frac{1}{2} \chi \left[ n_- (n_+)^{1/2} + (N-n_-) (N-n_+)^{1/2} \right] \sim \frac{1}{2} \chi n_- \left[ N^{1/2} + (n_+)^{1/2} - (N-n_+)^{1/2} \right] \, ,
\end{equation}
up to single-resonator terms (which lead to single-qudit phases). This interaction is closer to the cross-Kerr interaction in that the term in brackets is close to the line~$2 n_+ /\sqrt{N}$. 

Subsequent iterations apply permutations on smaller numbers of states (specified below) and produce effective Hamiltonians of the form
\begin{equation}
\mathcal{H}_{j} \sim \chi n_- f_j(n_+) \, ,
\end{equation}
with discrete functions~$f_j(n)$. An example for the case~$N=9$ (qudit dimension~$d=N+1 = 10$) is shown in Fig.~\ref{pfunctions}(a) for several iterations~$j=1 \to 4$. By choosing the weights~$\lambda_j$ appropriately, a linear combination of these functions can be made to be as linear as one desires. An example of such a combination for~$N=9$ is shown in Fig.~\ref{pfunctions}(b).  

\begin{figure}
\includegraphics[width=6in]{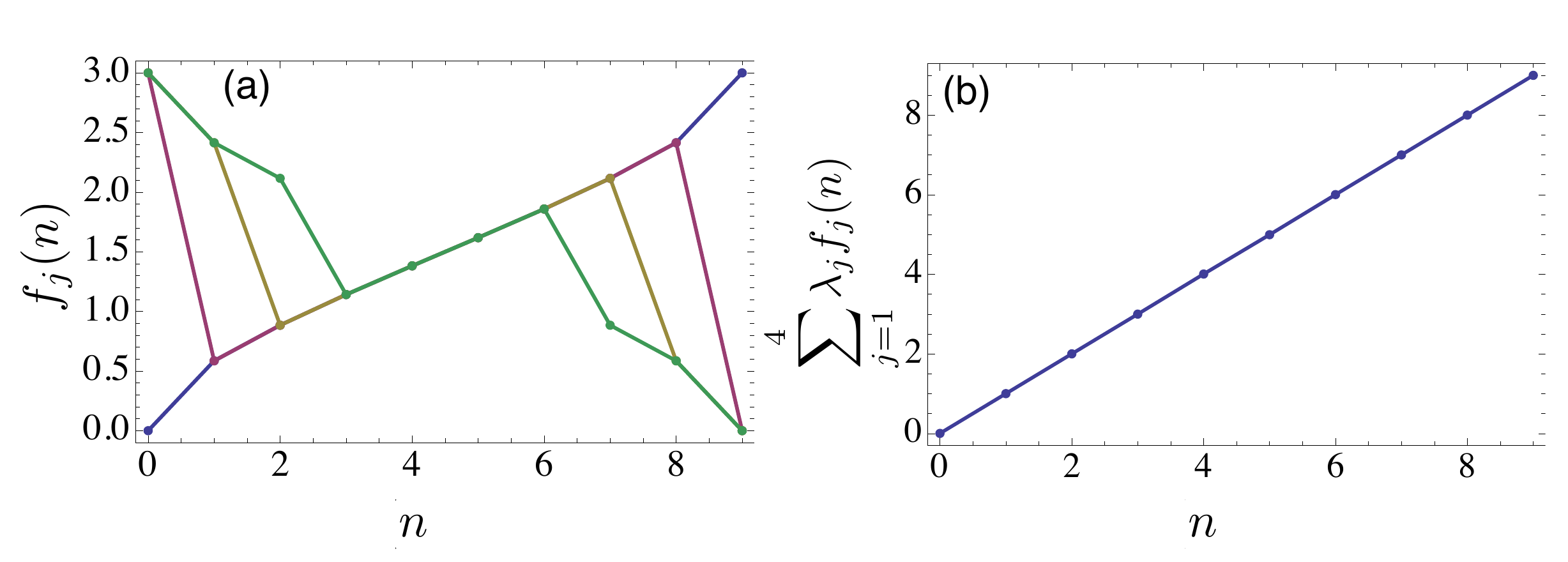}
\caption{(a) The discrete functions~$f_j(n)$ as a function of photon number~$n$ for step~$j=1, 2, 3, \ \mbox{and} \ 4$ of the permutation circuit (in blue, purple, yellow, and green; see text). (b) A linear combination of the functions~$f_j(n)$ with weights~$\lambda_1 = 1.7967$, $\lambda_2 = 0.2071$, $\lambda_3 = 0.0579$, and $\lambda_4 = 0.0317$ (and overall shift of $4.5$), chosen to approximate a linear function.}
\label{pfunctions}
\end{figure}

The specific permutations are chosen to be~$\mathcal{P}_{j,1} = \mathcal{I} \otimes \mathcal{P}_j$ and $\mathcal{P}_{j,2} = \mathcal{P}_1 \otimes \left( \mathcal{P}_1 \mathcal{P}_j \right)$, where
\begin{equation}
\mathcal{P}_2 |j\rangle = \left\{ \begin{array}{ll} 
|N\rangle & \mbox{if} \ j=0, \\
|0\rangle & \mbox{if} \ j=N, \\
|j\rangle &  \ \mbox{otherwise} ,
\end{array} \right.
\end{equation}
\begin{equation}
\mathcal{P}_3 |j\rangle = \left\{ \begin{array}{ll} 
|N\rangle & \mbox{if} \ j=0, \\
|N-1\rangle & \mbox{if} \ j=1, \\
|1\rangle & \mbox{if} \ j=N-1, \\
|0\rangle & \mbox{if} \ j=N, \\
|j\rangle & \ \mbox{otherwise} ,
\end{array} \right .
\end{equation}
and
\begin{equation}
\mathcal{P}_4 |j\rangle = \left\{ \begin{array}{ll} 
|N\rangle & \mbox{if} \ j=0, \\
|N-1\rangle & \mbox{if} \ j=1, \\
|N-2\rangle & \mbox{if} \ j=2, \\
|2\rangle & \mbox{if} \ j=N-2, \\
|1\rangle & \mbox{if} \ j=N-1, \\
|0\rangle & \mbox{if} \ j=N, \\
|j\rangle & \ \mbox{otherwise} .
\end{array} \right .
\end{equation}

\subsection{B. Gate Fidelities}

Using the permutation circuit, we analyze the gates using the process fidelity
\begin{equation}
\mathcal{F} = \frac{1}{d^4} \left| \mbox{trace}(U^{\dagger}_{\subs{ideal}} U)) \right|^2 \, ,
\end{equation}
where~$U = e^{-i \mathcal{H}_{\subs{total}} t}$ are $d^2 \times d^2$ matrices, single-qudit phases are removed, and the interaction time and weights are numerically optimized. The results of these calculations for the qudit \textsc{controlled-phase} gate (with $\theta_d = 2\pi/d$) are shown in Fig.~5 of the main text.

\end{widetext}

\end{document}